%% file: sausageRev.tex
\documentclass[12pt]{iopart}

\usepackage{iopams}

\usepackage{aas_macros}

\usepackage{pdflscape}

\newcommand{\jpa}{{\it J. Phys. A: Math. Theor.} }

\usepackage[colorlinks=true,citecolor=darkgreen]{hyperref}

\usepackage{graphicx}        
\usepackage{amssymb}         
\usepackage[usenames,dvipsnames]{color}          
\definecolor{darkblue}{rgb}{0,0,0.5}
\definecolor{darkgreen}{rgb}{0,0.6,0}
\definecolor{light-gray}{gray}{0.7}
\usepackage{epstopdf}
\usepackage{dsfont}

\DeclareFontFamily{OT1}{pzc}{}
\DeclareFontShape{OT1}{pzc}{m}{it}%
             {<-> s * [1.1500] pzcmi7t}{}
\DeclareMathAlphabet{\mathscr}{OT1}{pzc}%
                                 {m}{it}

\expandafter\let\csname equation*\endcsname\relax
\expandafter\let\csname endequation*\endcsname\relax
\usepackage{amsmath}

\newcommand{\fract}[2]{\leavevmode\kern.1em
          \raise.5ex\hbox{\the\scriptfont0 #1}\kern-.1em
    \raise.15ex\hbox{\the\scriptfont0 /}\kern-.08em\lower.25ex\hbox{\the\scriptfont0 #2}}

\newcommand{\half}{{\textstyle\frac{1}{2}}}

\newcommand{\rd}{\mathrm{d}}

\newcommand{\deriv}[2]{\frac{\rd#1}{\rd#2}}

\renewcommand{\L}{{\ell}}

\newcommand{\diag}{\mathop{\rm diag}}

\newcommand{\ri}{{i}}

  \renewcommand{\le}{\leqslant}

\begin{document}


\title[Sensitivity of Coronal Loop Sausage Mode Frequencies]{Sensitivity of Coronal Loop Sausage Mode Frequencies and Decay Rates to Radial and Longitudinal Density Inhomogeneities:\\ A Spectral Approach}

\date{\today}


\author{Paul S.~Cally$^1$ and Ming Xiong$^2$}
\address{$^1$School of Mathematical Sciences and Monash Centre for Astrophysics,\\ Monash University, Clayton, Victoria 3800, Australia}
\address{$^2$State Key Lab of Space Weather,
National Space Science Center,\\
Chinese Academy of Sciences,
Haidian District,
Beijing, China}
\ead{paul.cally@monash.edu, mxiong@spaceweather.ac.cn}


\begin{abstract}\noindent
Fast sausage modes in solar magnetic coronal loops are only fully contained in unrealistically short dense loops. Otherwise they are leaky, losing energy to their surrounds as outgoing waves. This causes any oscillation to decay exponentially in time. Simultaneous observations of both period and decay rate therefore reveal the eigenfrequency of the observed mode, and potentially insight into the tubes' nonuniform internal structure. In this article, a global spectral description of the oscillations is presented that results in an implicit matrix eigenvalue equation where the eigenvalues are associated predominantly with the diagonal terms of the matrix. The off-diagonal terms vanish identically if the tube is uniform. A linearized perturbation approach, applied with respect to a uniform reference model, is developed that makes the eigenvalues explicit. The implicit eigenvalue problem is easily solved numerically though, and it is shown that knowledge of the real and imaginary parts of the eigenfrequency is sufficient to determine the width and density contrast of a boundary layer over which the tubes' enhanced internal densities drop to ambient values. Linearized density kernels are developed that show sensitivity only to the extreme outside of the loops for radial fundamental modes, especially for small density enhancements, with no sensitivity to the core. Higher radial harmonics do show some internal sensitivity, but these will be more difficult to observe. Only kink modes are sensitive to the tube centres. {Variation in internal and external Alfv\'en speed along the loop is shown to have little effect on the fundamental dimensionless eigenfrequency, though the associated eigenfunction becomes more compact at the loop apex as stratification increases, or may even displace from the apex.}
\end{abstract}


\submitto{\jpa}

\maketitle


\section{Introduction}
Coronal magneto-seismology seeks to infer the physical characteristics of solar coronal structures, particularly coronal loops, from observations of low frequency magneto\-hydro\-dynamic (MHD) waves and oscillations. 

Currently unmeasurable physical parameters inside coronal loops such as magnetic field strength, field-aligned flow magnitude, plasma temperature, and effective adiabatic index can be probed by various coronal seismology inversion schemes.  Collective wave modes supported by magnetized tubes include kink and sausage modes. Both kink and sausage modes abound in the lower solar atmosphere.

Although much attention has been devoted to kink modes ($m_\text{c}=\pm1$, assuming an $\rme^{\ri m_\text{c}\theta}$ dependence on cylindrical angle $\theta$), there is a long history of interest in fast sausage modes ($m_\text{c}=0$) as well \cite{AscNakMel04aa,KopMelSte07aa,IngvanBra09aa,PasNakArb09aa,CheLiXio15aa,GuoCheLi16aa}. Sausage modes are better suited to probing the radial structure of loop filaments, whereas kink modes are more suited to exploring their cores. They are widely thought to be implicated in quasi-periodic pulsations (QPPs) \cite{NakMel09aa} in flare loops.

Slow sausage modes have independently received considerable attention \cite{De-IreWal02aa,NakVer05aa,De-09aa}. These  ``ankle-biter modes'' are essentially field-guided acoustic waves largely restricted to the lower reaches of loops. They are not present under the pressureless assumption adopted here, and will not be discussed further. Fast sausage waves on the other hand are fast magneto\-hydro\-dynamic waves partially (leaky) or totally (non-leaky) confined to over-dense magnetic fluxtubes acting as wave guides. 

Non-ideal MHD mechanisms such as electron heat conduction, ion viscosity, and finite plasma-$\beta$ are are not included at this stage as they are believed to be too weak to cause the temporal damping observed in QPP events \cite{KopMelSte07aa,IngvanBra09aa}. A detailed survey of the more extensive literature on coronal sausage waves and QPPs is given in \cite{CheLiXio15aa}.

Except for very short fat loops, (fast) sausage modes in coronal conditions are normally leaky. That is, they couple to radially outgoing oscillations in the surrounding plasma that extract energy from the loop oscillations \cite{Cal85aa,Cal86aa}. This causes them to exhibit complex eigenfrequencies, with negative imaginary parts. Observations of both period and decay rate therefore have the potential to usefully constrain loop properties.

Recently, \cite{CheLiXio15aa} and \cite{GuoCheLi16aa} explored sausage oscillations in pressureless straight tube models using numerical solution of the governing ordinary differential equation (ODE) in radius $r$, implementing a variety of density cross-sections. Finite plasma-$\beta$ is addressed in  \cite{CheLiXio16aa} but will not be considered here. These studies aimed to constrain the internal to external density contrast $\rho_\text{i}/\rho_\text{e}$ and the transverse Alfv\'en crossing time $R/a_\text{i}$, where $R$ is the loop radius and $a_\text{i}$ the internal Alfv\'en speed, as well as obtain some information on the steepness of the radial density profile in the boundary layer between loop and external corona. 

In this article, we address essentially the same model, though optionally with stratification along the loop allowed as well (the effects of stratification on kink waves have been well-studied: \cite{AndGooHol05aa,ErdVer07aa}, etc.). However, rather than using a direct numerical ODE solver, we adopt a spectral decomposition that has several desirable properties. In particular, the resulting matrix eigenvalue problem is highly diagonally dominant, with the off-diagonal terms representing the radial inhomogeneity. This makes it easy to linearize the problem (in essentially boundary layer width), and thereby to identify density inversion kernels that explicitly show which modes are sensitive to which parts of the loop. This is carried out for the unstratified case, where a consequent perturbation analysis yields an explicit formula for frequency variance in terms of radial inhomogeneity parameters. Accuracy of the linearized perturbation formula is tested against full numerical solution.

Finally, the method is extended to allow for stratification along the loop, both internally and externally (independently). It is found that even mild stratification can significantly affect damping rate in particular, especially for an unstratified (hot) loop in a stratified atmosphere.

\section{Mathematical Development}
\subsection{General Equations}
Consider an axisymmetric Alfv\'en speed distribution $a(r,z)$ in a pressureless straight ``loop'' of length $\L$ and radius $R$, surrounded by an external atmosphere with Alfv\'en speed dependent only on $z$, $a(r,z)=a_0(z)$ ($r>R$). In the pressureless plasma approximation, sound speed is neglected compared to Alfv\'en speed, freezing out slow modes. (Loops of non-uniform cross section are considered by \cite{PasNakArb09aa}, and non-zero plasma-$\beta$ is addressed by \cite{CheLiXio16aa}.)

\begin{figure}[htbp]
\begin{center}
\includegraphics[width=.45\textwidth]{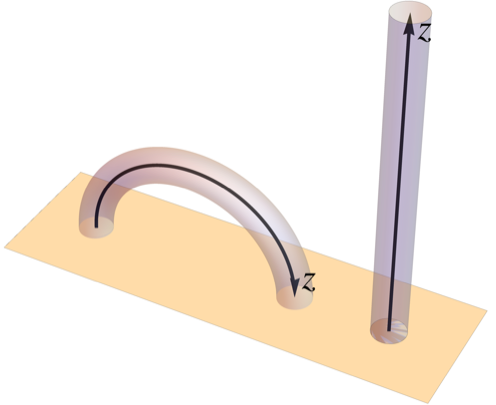}
\caption{ Schematic of a coronal loop (left) and the rectified flux tube model used in the analysis (right).}
\label{Fig:Loop}
\end{center}
\end{figure}

{ As is conventional, the loop is straightened for mathematical convenience \cite{RudRob02aa}, as depicted in Fig.~\ref{Fig:Loop}, thereby neglecting curvature effects \cite{SmiRobOli97aa,VanDebAnd04aa,VerFouNak06aa}. The coordinate $z$ is distance along the loop from one footpoint. The terms ``loop'' and ``tube'' will be used interchangeably throughout, despite the rectification.
}

If wave perturbations on this system are also assumed axisymmetric -- the sausage modes -- the linearized wave equation for the radial displacement $\xi$ assuming an $\exp(-i\omega t)$ time dependence is
\begin{equation}
\partial_r\left(\frac{1}{r}\,\partial_r\left(r\,\xi\right)\right)+\left(\partial_z^2+\frac{\omega^2}{a^2}\right)\xi=0,
\label{wave eqn}
\end{equation}
where $\partial_r$ represents the partial derivative with respect to $r$, etc. (see for example \cite{CheLiXio15aa}, Equation (6)). { The Alfv\'en speed $a$ is related to the magnetic field strength $B$ and plasma density $\rho$ via (in SI units)
\begin{equation}
a^2=\frac{B^2}{\mu\,\rho},
\end{equation}
where $\mu=4\pi\times 10^{-7}$ henry $\textrm{m}^{-1}$ is the permeability of free space \cite{Pri82aa}.}

For this sausage mode, there is no coupling to the Alfv\'en wave, unlike the case of kink waves, where coupling with the Alfv\'en wave results in resonant absorption and consequent kink mode decay. The wave under consideration therefore is a pure fast sausage wave.

The adopted boundary conditions are $\xi=0$ on $r=0$, $z=0$, and $z=\L$, and that $\xi$ and $\partial_r\xi$ match continuously to an outgoing or evanescent external solution at $r=R$.

Using separation of variables, the external solution may be written as
\begin{equation}
\xi_\text{e}(r,z)=\sum_{n=0}^\infty c_n H_1^{(1)}(L_n r) \, w_n(z),    \label{xiext}
\end{equation}
for arbitrary $c_n$, where $H_1^{(1)}$ is the Hankel function of the first kind and order 1 and $(L_n,\,w_n)$ is the $n^\text{th}$ eigenvalue/eigenfunction of the {regular} Sturm-Liouville equation
\begin{equation}
w_n'' +\left(\frac{\omega^2}{a_0(z)^2}-L_n^2\right)w_n=0   \label{SL}
\end{equation}
with boundary conditions $w_n(0)=0=w_n(\L)$. The $w_n$ will be alternately even and odd about $z=\L/2$ since $a_0(z)$ is assumed even. By general Sturm-Liouville theory, orthogonality {$\int_0^\L w_n\, w_\nu\,dz=\half\,\delta_{n\nu}$ (adopting a particular normalization)} is guaranteed provided $L_n^2\ne L_\nu^2$, making the $w_n$ convenient expansion functions.

Internally, a Bessel/Sturm-Liouville expansion of the radial displacement is adopted,
\begin{equation}
\xi(r,{z}) = \sum_{m=1}^\infty \sum_{n=1}^\infty Y_{mn}\, X_{mn}(r,z), \qquad (r\le R),\label{expand}
\end{equation}
where the orthogonal expansion functions are
\begin{equation}
X_{mn}=W_{mn}^{-1/2}J_1(l_{mn}r)w_n(z),   \label{X}
\end{equation}
and the $l_{m\,n}$ are the internal radial wavenumbers belonging to a radially uniform reference tube.
Here
\begin{equation}
W_{mn} =\L\left[\frac{R^2}{4} \left( J_0( l_{mn}R){}^2+ J_1( l_{mn}R){}^2\right)
     -\frac{R J_1(l_{mn}R) J_0(l_{mn}R)}{2 l_{mn}}\right]  \label{W}
\end{equation}
normalizes the kernel $\mathcal{K}_{mnpq}(r,z)=r\,X_{mn}(r,z)X_{pq}(r,z)$:
\begin{equation}
\int_0^\L\!\!\! \int_0^R  \mathcal{K}_{mnpq}(r,z)\, dr \,dz= \delta_{mp} \delta_{n q},
\label{orthog}
\end{equation}
where $\delta_{mn}$ is the Kronecker delta. 
The orthogonality (\ref{orthog}) is valid whether $l_{mn}$ is real or complex.

If the external medium is not stratified in $z$, i.e., $a_0=\text{constant}$, then $w_n(z)=\sin k_n z$, where $k_n=n\pi/\L$ are the longitudinal wavenumbers. The expansion is then Bessel/Fourier. In that case, the $L_n$ are given simply by
\begin{equation}
L_n = \left(\frac{\omega^2}{a_0^2}-k_n^2\right)^{1/2},   \label{L}
\end{equation}
$-\pi/2<\arg L_n\le\pi/2$. Otherwise, they are the eigenvalues of Equation (\ref{SL}), as above.

The internal radial wavenumbers $l_{mn}$ are determined by the condition that each of the expansion functions individually couples to an evanescent or outgoing radiation solution in $r>R$, i.e., that both $\xi$ and $\partial_r\xi$ are continuous at $R$. That is,
\begin{equation}
l_{mn}\frac{J_0\left(l_{mn}R\right)}{J_1\left(l_{mn}R\right)} = L_{n}\frac{H^{(1)}_0\left(L_{n}R\right)}{H^{(1)}_1\left(L_{n}R\right)} .\label{JHmatch}
\end{equation}

If $\omega\in\mathds{R} $ and $\omega^2<a_0^2k_n^2$ then $L_n$ is pure imaginary ($\arg L_n=\pi/2$), the Hankel functions become modified Bessel functions of the second kind (the evanescent exterior), and the $l_{mn}$ are real.

It is convenient to define $Q(r,z)=a(r,z)^{-2}-a_0(z)^{-2}$. Multiplying Equation (\ref{wave eqn}) by $r X_{mn}$ and integrating yields
\begin{equation}
\sum_{p=1}^\infty\sum_{q=1}^\infty Y_{pq}V_{mnpq} +\omega^{-2}\left(L_n^2-l_{mn}^2\right)Y_{mn} =0, \label{Yeqn}
\end{equation}
where
\begin{equation}
V_{mnpq} = \int_0^\L\!\!\! \int_0^R  Q(r,z)\,\mathcal{K}_{mnpq}(r,z)\,dr\,dz.  \label{V}
\end{equation}
When truncated ($1\le m,p\le M$ and $1\le n,q\le N$), Equation (\ref{Yeqn}) has non-trivial solutions if and only if the $MN\times MN$ matrix 
\begin{equation}
 \omega^{-2}\diag\left[L_n^2-l_{mn}^2\right]+V =K\,A\,K \label{singular}
\end{equation}
is singular, where $V$ has components $V_{mnpq}$. Here $K^2=\omega^2 I-\diag\left[L_n^2-l_{mn}^2\right]$, and Equation (\ref{singular}) defines $A=K^{-2}+K^{-1}V K^{-1}-\omega^{-2} I$. In practice, the eigenvalue problem is solved iteratively to make $A$ singular.

\subsection{Uniform Density Tube}
In the unstratified case, $a_0=\text{constant}$ and Equation (\ref{L}) applies, it is convenient to define $Q=\Delta/a_0^2$, where $\Delta$ is the fractional density increase inside the tube. 

Equation (\ref{JHmatch}) is also the appropriate matching equation when $\Delta$ is uniform and there is an Alfv\'en speed discontinuity at $R$. In that case though, the $X_{mn}$ are not just expansion functions; they are independent modes with their own eigenfrequencies  
\begin{equation}
\omega_{mn} = a\left(k_n^2+l_{mn}^2\right)^{1/2} = \frac{a_0}{\sqrt{1+\Delta}}\left(k_n^2+l_{mn}^2\right)^{1/2}.  \label{dispUniform}
\end{equation}

The condition for a sausage mode to be trapped (non-leaky) is that $L_n$ be imaginary, i.e., $k_n^2>\omega^2/a_0^2=(k_n^2+l_{mn}^2)/(1+\Delta)$, hence $k_n^2\Delta>l_{mn}^2$, or $\Delta > l_{mn}^2\L^2/(n^2\pi^2)$. With $l_{mn}=\mathcal{O}(j_{1,m}/R)$, trapped sausage modes are only to be expected for very large density enhancements $\Delta\gtrsim (j_{1,m}^2/n^2\pi^2)(\L^2/R^2)$, where $j_{1,m}=\mathcal{O}(m\pi)$ as $m\to\infty$ is the $m^\text{th}$ zero of the $J_1$ Bessel function. For typical long thin loops this is highly implausible for the longitudinal fundamental $n=1$ or low harmonics, even for $m=1$, so only leaky sausage modes need be considered. (An example of a short dense uniform loop that supports a non-leaky sausage mode is $R=0.5$, $\L=a_0=1$,  $\Delta=5>\L^2/R^2$ where $\omega=2.506$ for $m=n=1$, for which $\omega<a_0k_1=\pi$ as anticipated.)

The eigenfrequencies set out in Equation (\ref{dispUniform}) provide useful reference states for perturbative solution of the non-uniform $\Delta$ case. Table \ref{tab:uniform} lists the first few uniform-density eigenfrequencies (all leaky) for $\Delta=0.2$, 1, and 4 for fat ($R=0.1$) and thin ($R=0.02$) loops. The $n$ corresponds to the longitudinal wavenumber $k_n$, and the $m$ refers to the radial order. These are the so-called ``Trig Modes'' analysed by \cite{Cal85aa,Cal86aa}. Without loss of generality the loop length and external Alfv\'en speed have both been scaled to unity.

\begin{landscape}
\begin{table}
\caption{Eigenfrequencies $\omega$ for uniform $\Delta=0.2$, 1, and 4, with $\L=a_0=1$ for two loop radii, $R=0.1$ and $R=0.02$. \label{tab:uniform} }
\small
\begin{tabular}{ccccccc}
\br
& \multicolumn{3}{c}{Thick Loop $R=0.1$} & \multicolumn{3}{c}{Thin Loop $R=0.02$} \\ 
& \multicolumn{3}{c}{\hrulefill} & \multicolumn{3}{c}{\hrulefill} \\
$\Delta=0.2$&{$n=1$} & {$n=2$} & {$n=3$} & {$n=1$} & {$n=2$} & {$n=3$}
\\
\br
$m=1$ &$20.1946 -14.2687 i$ & $20.3986 -13.9122 i$ & $20.7507 -13.3102 i $ & $100.651 -71.9105 i$ & $100.692 -71.8397 i$ & $100.758 -71.7218 i$\\
$m=2$ & $49.6389 -14.1156 i$ & $49.8441 -14.0108 i$ & $50.1861 -13.8384 i$ & $247.866 -70.7466 i$ & $247.907 -70.7255 i$ & $247.976 -70.6903 i$\\
$m=3$ & $78.5200 -14.1047 i$ & $78.6654 -14.0592 i$ & $78.9073 -13.9839 i$ & $392.367 -70.5964 i$ & $392.397 -70.5872 i$ & $392.445 -70.5720 i$\\
\mr
$\Delta=1$\\
\mr
$m=1$ & $15.6463 - 5.89325 i$ & $ 15.8805 - 5.49913 i$ & $ 16.3094 - 4.82267 i$ & $77.8722 -30.0872 i $ & $77.9167 -30.0098 i$ & $77.9911 -29.8807 i$\\
$m=2$ & $ 38.4549 - 6.16217 i$ & $ 38.6271 - 6.08495 i$ & $ 38.9135 - 5.95805 i$& $191.999 -30.9353 i$ & $192.033 -30.9197 i$ & $192.091 -30.8937 i$\\
$m=3$ & $ 60.8228 - 6.20317 i$ & $ 60.9392 - 6.17138 i$ & $ 61.1328 - 6.11875 i$& $303.928 -31.0669 i$ & $303.951 -31.0605 i$ & $303.99 -31.0499 i$  \\
\mr
$\Delta=4$\\
\mr
$m=1$ & $10.1053 -1.68047 i$ & $10.3673 -1.29651 i$ & $10.9546 - 0.62556 i$ &  $50.156 -\ \,8.9945 i$ & $50.201 -8.92111 i$ & $50.2771 -8.79836 i$       \\
$m=2$  & $24.3446 -2.04559 i$ & $24.4581 -1.98371 i$ & $24.6469 -1.88016 i$ & $121.542 -10.3269 i$ & $121.564 -10.3145 i$ & $121.602 -10.2939 i$  \\
$m=3$  & $38.4745 -2.10672 i$ & $38.5491 -2.08220 i$ & $38.6733 -2.04135 i$ & $192.253 -10.5728 i $ & $192.268 -10.5679 i $ & $192.293 -10.5597 i$\\
\br
\end{tabular}
\end{table}%
\end{landscape}


Clearly, dependence of frequency on $n$ is weak, and on $m$ is strong, indicating the relative dependence of frequency on the longitudinal and radial variations in $\xi$. Sausage modes, by their nature, are sustained by radial total pressure variations, not longitudinal tension (which is the primary mechanism for kink waves).

\subsection{Nondimensionalization}
For numerical purposes it is convenient to nondimensionalize throughout. Adopting as the fiducial length $\L$ (the loop length) and velocity $a_\text{e}(\L/2)$ (the external Alfv\'en speed at the loop apex), we also have the time scale $\tau=\L/a_\text{e}(\L/2)$. Hence, any dimensionless frequency $\omega$ calculated below should be understood as corresponding to the dimensional frequency $\omega/\tau$. The radii are of course all relative to $\L$.

\section{Unstratified Loop with Boundary Layer}
In this section, it is assumed that both the loop and background are unstratified in $z$. Then  $Q(r,z)=\Delta(r,z)/a_0^2$, where $\Delta$ is the fractional density increase inside the tube with respect to the exterior.

\subsection{Sensitivity of Eigenfrequencies to Density Inhomogeneities}

Assume that the spectral expansion is truncated to $m\le M$ and $n\le N$. Therefore, Equation (\ref{Yeqn}) represents $MN$ equations in $MN$ unknowns, $Y_{mn}$. Let $x_{(m-1)N+n}=Y_{mn}$ be a single vector rearrangement: $x_1=K_{1\,1}Y_{1\,1}$, $x_2=K_{1\,2}Y_{1\,2}$, \ldots, $x_N=K_{1\,N}Y_{1\,N}$, $x_{N+1}=K_{2\,1}Y_{2\,1}$, \ldots, where $K_{mn}=(l_{mn}^2+k_n^2)^{1/2}$. Then the truncated Equation (\ref{Yeqn}) may be written in matrix form
\begin{equation}
Bx=\lambda x  \label{Bx}
\end{equation}
where $B=K^{-2}+K^{-1}U K^{-1}$ and $\lambda=a_0^2/\omega^2$. Here the diagonal matrix $K=\diag[K_{1\,1},\,K_{1\,2},\,\ldots,\, K_{1\,N},\, K_{2\,1},\,\ldots,K_{M\,N}]$, and $U$ is the $MN\times MN$ symmetric matrix with $(i,j)$ element $U_{mnpq}=\int_0^\L\int_0^R \Delta\,\,\mathcal{K}_{mnpq}\,dr\,dz=V_{mnpq}/a_0^2$, where { $m=\lfloor 1+(i-1)/N\rfloor$, $n=1+(i-1)\!\!\mod N$, $p=\lfloor 1+(j-1)/N\rfloor$, and $q=1+(j-1)\!\!\mod N$}, i.e.,
\begin{equation}
U=
\begin{pmatrix}
U_{1111}& U_{1112} &\ldots & \ldots &U_{111N} & U_{1121} & \ldots & U_{11MN}\\[4pt]
U_{1211} &U_{1212} & U_{1213} & \ldots & U_{121N} & U_{121N} & \ldots & U_{12MN} \\
\vdots & \ldots & \ddots & \ldots & \ldots & \ldots & \ldots & \vdots \\
\vdots & \ldots &  \ldots & \ddots & \ldots & \ldots & \ldots & \vdots \\
U_{1N11} & U_{1N12} & \ldots & \ldots & U_{1N1N} & \ldots & \ldots & U_{1NMN} \\
U_{2111} & U_{2112} & \ldots & \ldots & \ldots & \ddots & \ldots & \vdots \\
\vdots & \ldots & \ldots & \ldots & \ldots & \ldots & \ddots & \vdots \\
U_{MN11} & \ldots & \ldots & \ldots & U_{MN1N} & \ldots & \ldots &  U_{MNMN}
\end{pmatrix}.
\end{equation}

By Gerschgorin's Circle Theorem \cite{Ger31aa}, all eigenvalues $\lambda=a_0^2/\omega^2$ lie within the union of the Gerschgorin disks $D_{mn}$ in the complex $\lambda$-plane centred at
\begin{equation}
\bar\lambda_{mn} = \frac{1+U_{mnmn}}{K_{mn}^2},  \label{Lambda}
\end{equation}
with radius
\begin{equation}
R_{mn}=\mathop{\sum^\infty\sum^\infty}_{\substack{p=1\ q=1\\ (p,q)\ne(m,n)}} \left|\frac{U_{mnpq}}{K_{mn}K_{pq}}\right| .
\end{equation}
Furthermore, any set of $\nu$ contiguous disks that is disjoint from the other disks contains exactly $\nu$ eigenvalues. The disk centres $\bar\lambda_{mn}$ are in exact accord with Equation (\ref{dispUniform}), the uniform-tube eigenfrequencies.

Several points should be made about the eigenvalue equation (\ref{Bx}).
\begin{enumerate}
\item The off-diagonal terms in $U$ vanish for uniform $\Delta$, while the diagonal terms $U_{mnmn}$ are just $\Delta$.
\item The eigenvalues $\lambda$ also occur implicitly in $K$ and $U$, and hence in $B$, through the external dispersion relation (\ref{L}) via the radial wave numbers $l_{mn}$ defined by Equation (\ref{JHmatch}).
\item Increasing the truncation limits $M$ and $N$ does not alter the Gerschgorin centre {$\bar\lambda_{mn}$}, but does (slightly) increase the radius $R_{mn}$. Typically, for the fundamental mode $m=n=1$, introducing a boundary layer (with $\alpha=0.01$ as in Section \ref{resultsIndep} below) and increasing $M$ and $N$ from 3 to 4 results in a change in the numerically determined frequency $\omega_{mn}$ only in the sixth significant figure. For the $m=2$, $n=1$ first radial overtone, the change is in the fifth significant figure. In either case, $M=N=3$ is adequate in practice.
\end{enumerate}

With that understanding in place, a linearized perturbation approach offers useful insights.

\subsection{Perturbative Approach}
The Gerschgorin disk centres $\bar\lambda_{mn}$, as defined by Equation (\ref{Lambda}), provide estimates of the true eigenvalues, becoming exact as the off-diagonal terms vanish with diminishing inhomogeneity. However, since $\bar\lambda_{mn}$ actually depends on $\lambda$, the estimate is implicit rather than explicit. Assuming $|\lambda-\tilde\lambda_{mn}|$ is small though, where the tilde refers to the uniform reference state, a linear expansion may be made. Note that this does not require $\Delta-\tilde\Delta$ to be pointwise-small, only that $\int_0^\L\!\!\int_0^R  \Delta(r,z)\,\mathcal{K}_{mnpq}(r,z)\,dr\,dz\ll \tilde\Delta$ for all $m$, $n$, $p$, $q$. In particular, this is valid for a thin enough boundary layer in which $\Delta$ drops to zero on the edge, despite the total drop being large.

Recalling that $B$ is a function of $\lambda$, and expanding $\lambda=\tilde\lambda+\delta\lambda$, $B=\tilde B+\delta B$, and $x=\tilde x+\delta x$, Equation (\ref{Bx}) becomes $(\tilde B-\tilde\lambda\, I)\delta x+(\delta B-\delta\lambda\,I)x=0$ to first order in small quantities. If $\lambda=\lambda_{mn}$ say, then the $(m,n)$ row (i.e., the $((m-1)N+n)^\text{th}$ row) of $B-\lambda\, I$ is identically zero. Hence, since the associated eigenvector is $\tilde x=(0,0,\ldots,0,1,0,\ldots,0)^T$, where the ``1'' is the $(m,n)$ element, it follows that
\begin{equation}
\delta\lambda_{mn} = \delta B_{mnmn},
\end{equation}
where $\delta B_{mnmn}$ is the $(m,n)$ diagonal element. In other words, the perturbation to any eigenvalue is just the perturbation to the corresponding diagonal element of $B$, to leading order. That means that the perturbed eigenvalue is simply the centre $\bar\lambda_{mn}$ of the $D_{mn}$ Gerschgorin disk, to first order.

With that result, it follows from Equation (\ref{Lambda}) that
\begin{equation}
\omega_{mn}^2 = \tilde\omega_{mn}^2 + \frac{2a_0^2}{1+\tilde\Delta}\,\tilde l_{mn}\delta l_{mn}-\tilde\omega_{mn}^2\frac{U_{mnmn}-\tilde\Delta}{1+\tilde\Delta}  \label{omega pert}
\end{equation}
to first order. The departure of $U_{mnmn}$ from $\tilde\Delta$ is entirely due to the direct change in $\Delta$ and not the change in $l_{mn}$, since the normalization (\ref{orthog}) remains valid for any $l_{mn}$. Hence, $U_{mnmn}$ may be calculated using the $\tilde l_{mn}$ radial wavenumbers. The additional term proportional to $\delta l_{mn}$ results from the additional dependence of the radial wavenumbers $l_{mn}$ on $\omega$ via the matching condition (\ref{JHmatch}).

{
Now, 
\begin{equation}
\delta l_{mn} = \deriv{l_{mn}}{L_n}\,\delta L_n =  \deriv{l_{mn}}{L_n}\,\deriv{L_n}{\omega}\,\delta\omega = \frac{\omega}{a_0^2\,L_n}\deriv{l_{mn}}{L_n}\,\delta\omega = \frac{G_{mn}}{a_0^2\, l_{mn}}\,\omega\,\delta\omega, \label{delta l}
\end{equation}
defining the dimensionless quantity
\begin{equation}
\begin{split}
 G_{mn}&=\frac{l_{mn}}{L_n}\deriv{l_{mn}}{L_n}\\
 & = \frac{l_{mn}}{L_n}
 \frac{J_1(l_{mn}R)^2}{H_1^{(1)}(L_nR)^2}
 \frac{2H_0^{(1)}(L_nR)H_1^{(1)}(L_nR)-L_nR\left(H_1^{(1)}(L_nR)^2+H_0^{(1)}(L_nR)^2 \right)}
 {2J_0(l_{mn}R)J_1(l_{mn}R)-lR\left(J_1(l_{mn}R)^2+J_0(l_{mn}R)^2\right)},
 \end{split}
\end{equation}
which is obtained by differentiating Equation (\ref{JHmatch}). Ultimately, the fractional perturbation in the eigenfrequency of the $(m,n)$ mode is}
\begin{equation}
\frac{\delta\omega_{mn}}{\tilde\omega_{mn}} = -\frac{\delta U_{mnmn}}{2\left(1+\tilde\Delta
-G_{mn}\right)}  \label{delta omega}
\end{equation}
to first order, where
\begin{equation}
\delta U_{mnmn}=U_{mnmn}-\tilde\Delta  = \int_0^\L\!\!\! \int_0^R  \left(\Delta-\tilde\Delta\right)\,\mathcal{K}_{mnpq}(r,z)\,dr\,dz
\end{equation}
is the kernel-averaged variation in density from the uniform state. Note that $\tilde\Delta$ has not been assumed small, but all the $\delta$-terms have.

Typically, the corrections $G_{mn}$ are not negligible. For example, with $R=0.1$ and $a_0=\L=1$, $G_{1\,1}=0.87-0.25\,i$ for $\tilde\Delta=0.2$; $G_{1\,1}=0.71-0.68\,i$ for $\tilde\Delta=1$; and $G_{1\,1}=0.35-1.33\,i$ for $\tilde\Delta=4$.

An arbitrary density perturbation cannot uniquely be reconstructed from even a complete and perfect knowledge of all the sausage mode frequencies. Most obviously, $\delta U_{mnmn}$ is insensitive to any part of $\Delta(r,z)$ that is odd about $z=\L/2$, because $X_{mn}^2$ is even about that point for all $n$. Consequently, the odd part of $\Delta$ depends only, and very weakly, on the off-diagonal terms that enter at higher order.

\begin{figure}[htbp]
\begin{center}
\includegraphics[width=\hsize]{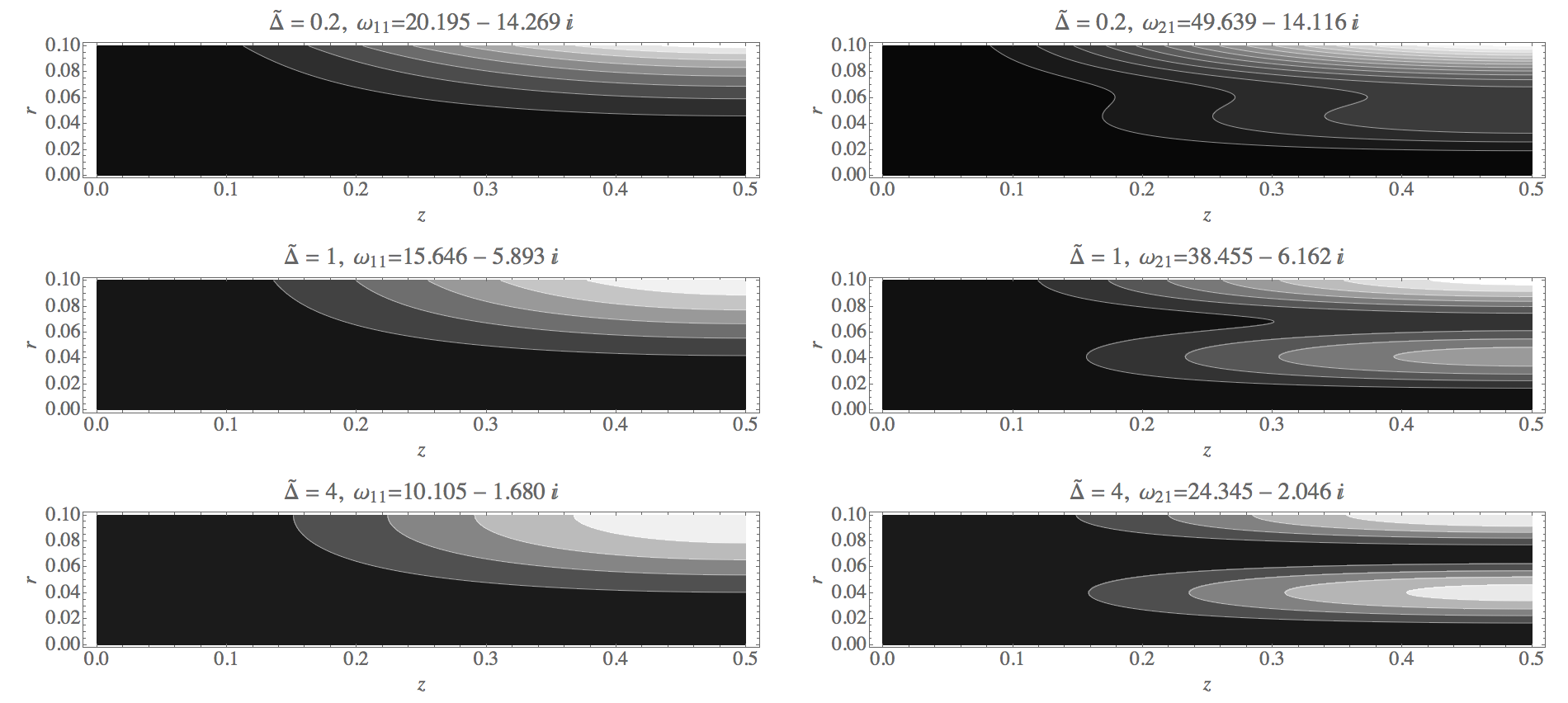}
\caption{ The absolute values of the sensitivity kernels, $|\mathcal{K}_{m1m1}(r,z)|$, for the thick loop $R=0.1$ with $m=1$ (left column) and $m=2$ (right column), with $\tilde\Delta=0.2$, 1, and 4 (top to bottom). The contours are 10, 20, 30, \ldots in all panels, but the shading is scaled to black for no sensitivity ($<10$) and white for maximal sensitivity separately in each panel in order to present maximum contrast. Only $0<z<\half$ is shown as these kernels are symmetric about $z=\half$. Note that each of these kernels is normalized to unity overall, $\int_0^\ell\int_0^R \mathcal{K}_{m1m1}(r,z)\,dr\,dz=1$.}
\label{fig:kernels}
\end{center}
\end{figure}

The density sensitivity kernels $\mathcal{K}_{mnmn}$ for the thick loop $R=0.1$ are displayed in Fig.~\ref{fig:kernels} for $m=1$ and 2 (the radial fundamental and first overtone), $n=1$ (longitudinal fundamental), and the three values of $\tilde\Delta$ listed in Table \ref{tab:uniform}. 

The radial fundamental $m=1$ clearly gives good sensitivity to density at the loop edge. The first harmonic is more sensitive to the interior, especially for larger $\tilde\Delta$. Because of the factor $r$ in $\mathcal{K}_{mnpq}$, there is no sensitivity at all to the tube centre. This is to be expected, since $r=0$ is a node for sausage modes. Indeed, it is a node for all higher azimuthal order modes except for kink modes, so only kink mode frequencies will be sensitive to the centre. Similarly, the loop apex is a node for $n=2$ {(not shown)}, or any even $n$. The $n=2$ modes are most sensitive at $z=\L/4$ and $3\L/4$.

Corresponding plots of the kernels for the thin loop $R=0.02$ are very similar (not shown). 

\subsection{Results: Density Independent of $z$}  \label{resultsIndep}

\begin{figure}
\begin{center}
\includegraphics[width=0.7\textwidth]{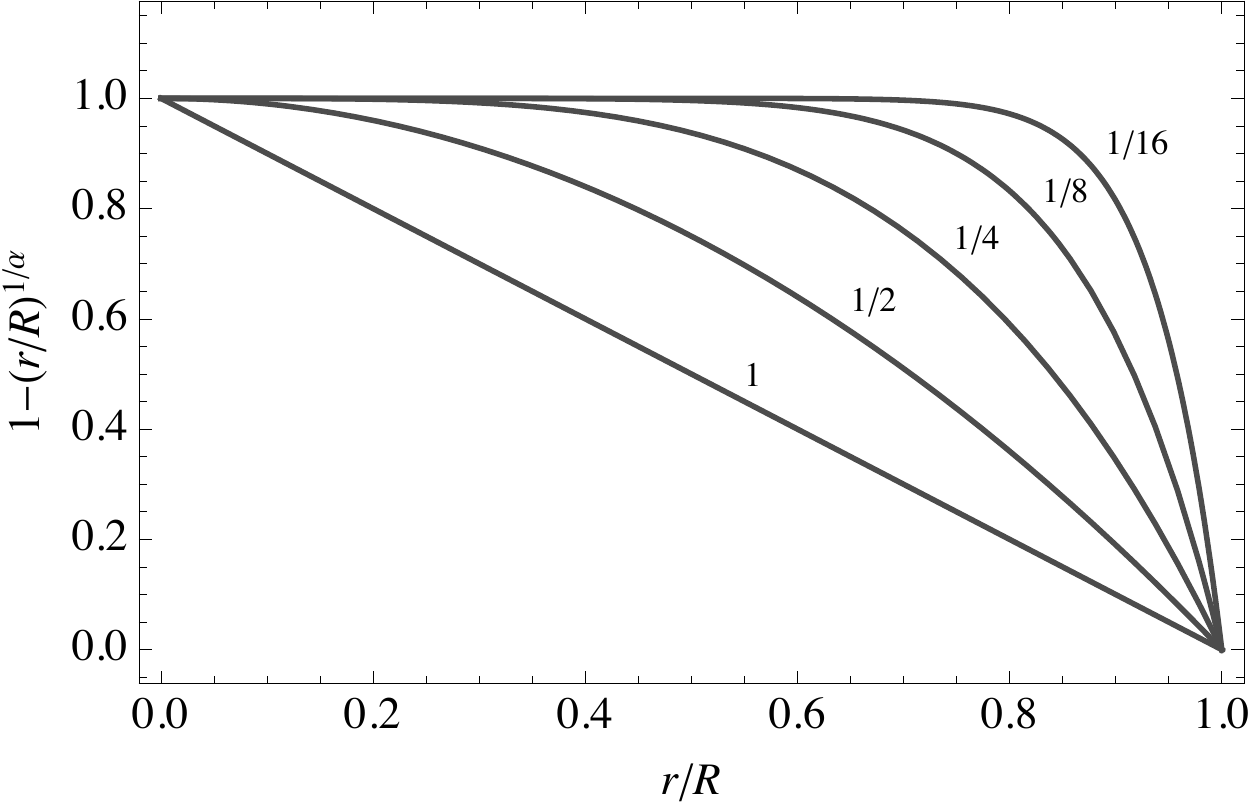}
\caption{Radial density perturbation model as given by Equation (\ref{Delta 1D}) 
for $\Delta_0=1$ and $\alpha=2^{-4},\,2^{-3},\ldots,\,1$ as labelled.}
\label{fig:BL}
\end{center}
\end{figure}

Consider a two-parameter density model independent of $z$,
\begin{equation}
{\Delta(r)} = \Delta_0\,\left(1-(r/R)^{1/\alpha}\right), \label{Delta 1D}
\end{equation}
that has a central plateau peaking at $\Delta_0$ at the centre, and a boundary layer falling continuously to zero at the edge $R$. The boundary layer becomes thinner as $\alpha$ diminishes (see Fig.~\ref{fig:BL}), with its thickness at half-height being $(1-2^{-\alpha})R$. The uniform reference model is recovered for $\Delta_0=\tilde\Delta$ and $\alpha\to0$.

\begin{figure}
\begin{center}
\includegraphics[width=0.95\textwidth]{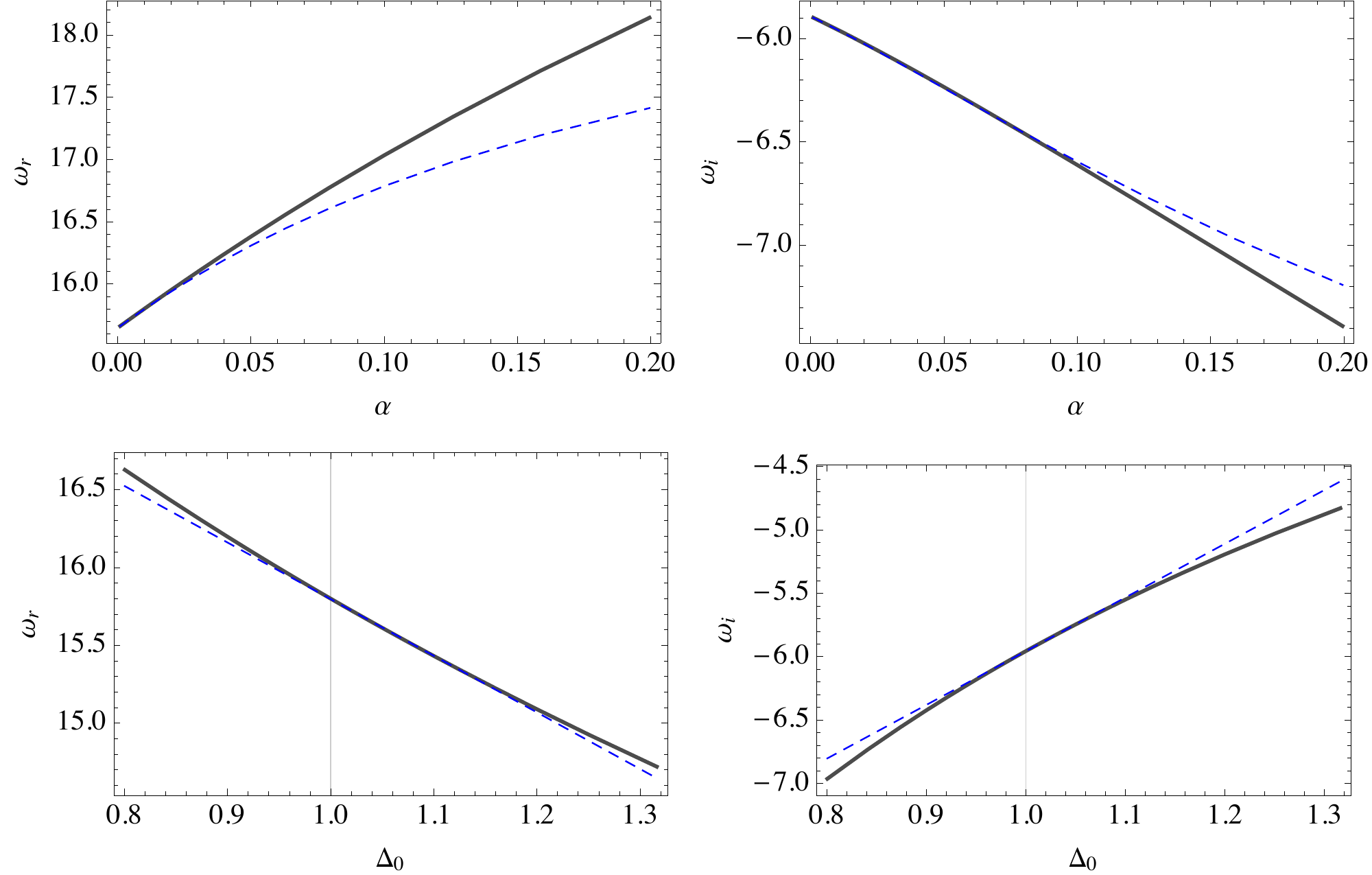}
\caption{Comparison of exact (full curves) and perturbation (dashed) frequencies $\omega_{1\,1}=\omega_r+\ri\,\omega_i$ for the case $R=0.1$, $\tilde\Delta=1$ with $M=N=4$. Top row: with $\Delta_0=1$ and variable $\alpha$. Bottom row: with $\alpha=0.01$ and variable $\Delta_0$. The real part of $\omega$ is depicted in the left column, and the imaginary part in the right column. The centre of the perturbation expansion, $\Delta_0=1$, is indicated by the vertical line in the lower panels.}
\label{fig:expert}
\end{center}
\end{figure}

\subsubsection{Linear Perturbation Solutions}

Figure \ref{fig:expert} compares the true (numerically derived) eigenfrequencies with those given by the perturbation formula (\ref{delta omega}) for variations around $\Delta_0=1$, $\alpha=0$, with $R=0.1$. It confirms the perturbation formula, and indicates its range of validity.

\subsubsection{Nonlinear Solution}

Although more expensive, the eigenfrequencies $\omega_{mn}$ for a specific model may be determined numerically without recourse to linearization. This is achieved using a standard root-finder to iteratively make $B-\lambda I$ singular (where $\lambda=a_0^2/\omega^2$; recall that $B$ also depends on $\lambda$ in a complicated manner) given $\Delta_0$ and $\alpha$. A starting guess is supplied by linear theory, allowing the selection of $m$ and $n$.

\begin{figure}[htbp]
\begin{center}
\includegraphics[width=.9\textwidth]{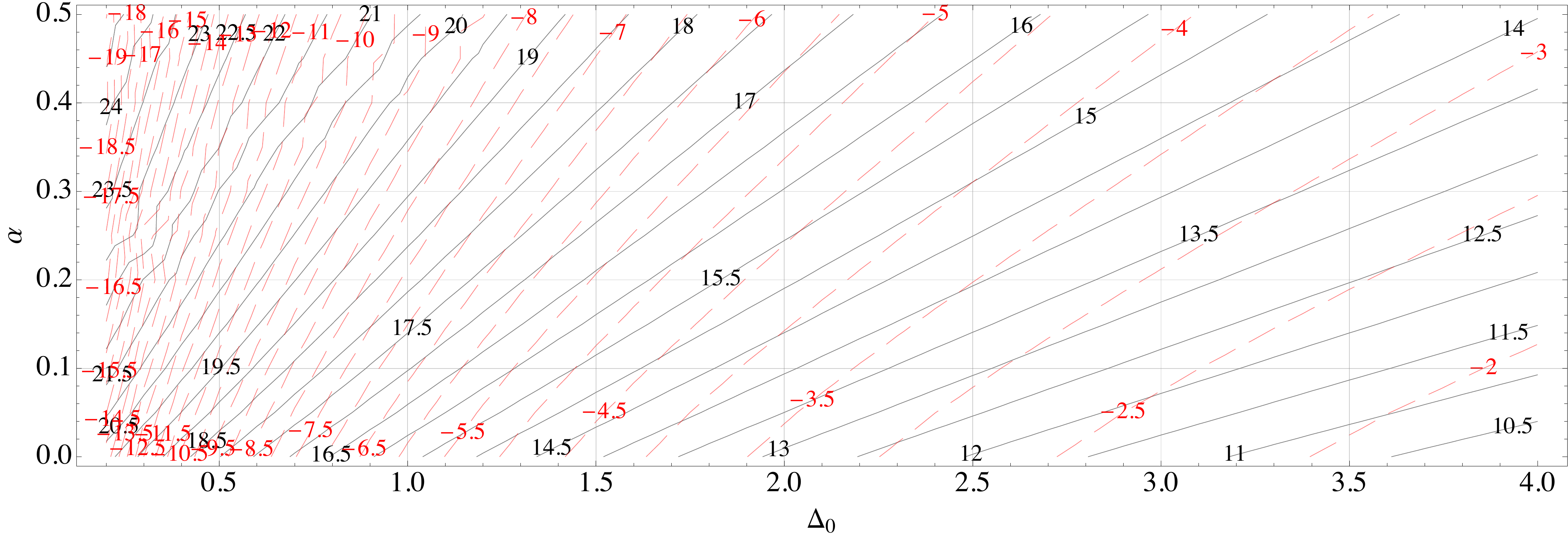}\\[6pt]
\includegraphics[width=.9\textwidth]{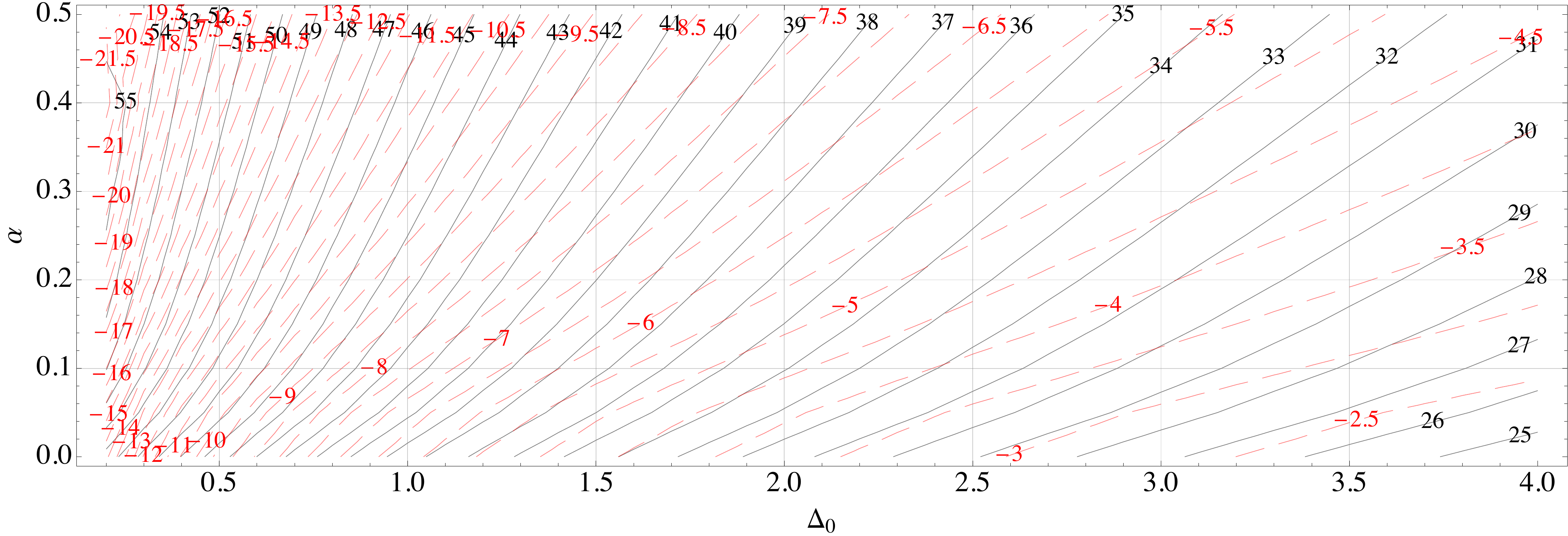}
\caption{Contours of the real (black) and imaginary (red dashed) parts of $\omega_{1\,1}$ (upper panel) and $\omega_{2\,1}$ (lower panel), obtained using non-linear iteration with $M=N=3$ against the two parameters $\Delta_0$ and $\alpha$ of the model of Equation (\ref{Delta 1D}) for $R=0.1$, $\L=a_0=1$.}
\label{fig:omegaNL}
\end{center}
\end{figure}

\begin{figure}[htbp]
\begin{center}
\includegraphics[width=.9\textwidth]{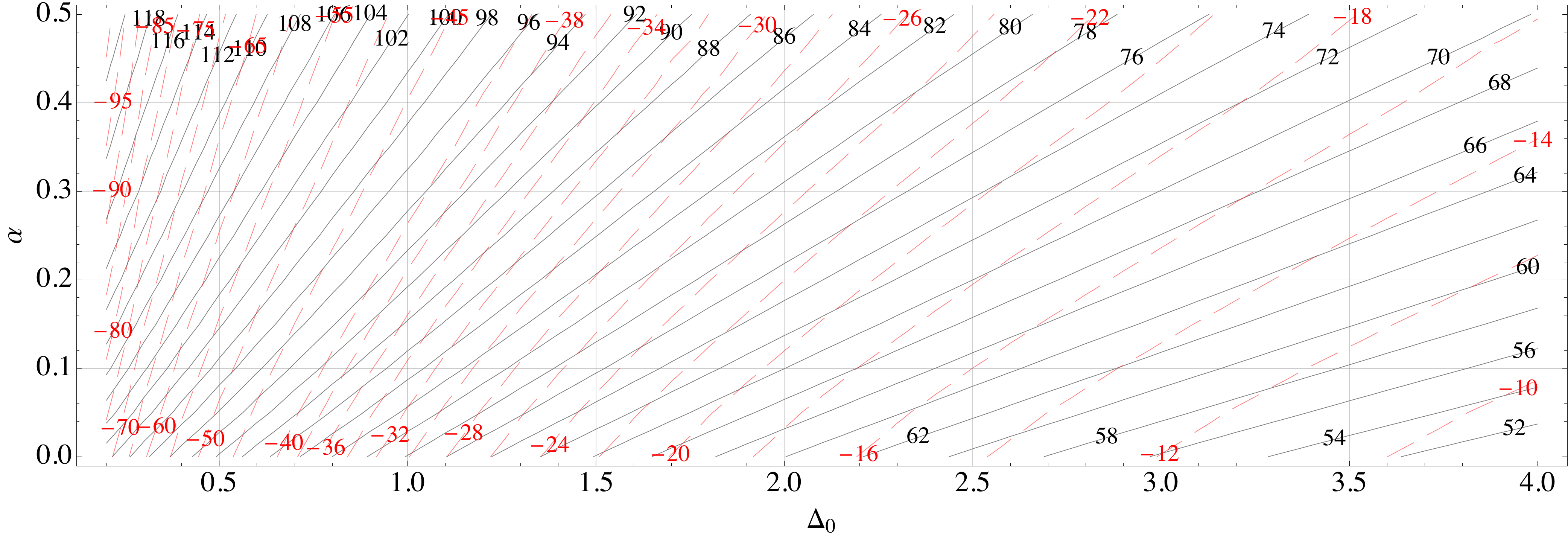}
\caption{Contours of the real (black) and imaginary (red dashed) parts of $\omega_{1\,1}$ obtained using non-linear iteration with $M=N=3$ against the two parameters $\Delta_0$ and $\alpha$ of the model of Equation (\ref{Delta 1D}) for the thin tube $R=0.02$, $\L=a_0=1$.}
\label{fig:omegaNLR02}
\end{center}
\end{figure}

Figure \ref{fig:omegaNL} shows contours of the real and imaginary parts of $\omega_{1\,1}$ and $\omega_{2\,1}$ for the thick tube $R=0.1$ over a wide range of $\Delta_0$ and $\alpha$. It was calculated using $M=N=3$ spectral resolution. Note that there is now no need for a reference model $\tilde\Delta$. Figure \ref{fig:omegaNLR02} similarly depicts $\omega_{1\,1}$ for the thin tube $R=0.02$.  

The fact that the contours of the real and imaginary parts of $\omega$ in Figures \ref{fig:omegaNL} and \ref{fig:omegaNLR02} for the most part cross at angles that are not too fine suggests that observations of period and decay rate are sufficient to determine $\Delta_0$ and $\alpha$ with some accuracy, assuming of course that the model (\ref{Delta 1D}) is valid. Simultaneous inversions using two modes, say the fundamental $\{1,1\}$ and first radial overtone $\{2,1\}$, would lend validity to the model if the resulting $(\Delta_0,\alpha)$ were consistent.

\section{Formulation with a Stratified Loop and Atmosphere}\label{Sec:strat}
We return to the gravitationally stratified atmosphere. For concreteness, assume that the external density diminishes exponentially with height in a semicircular loop, so that
\begin{equation}
a_0(z)^2 = a_1^2 \,\rme^{-\beta(1-\sin\pi z/\L)}  \label{a0strat}
\end{equation}
where $a_1$ is the external Alfv\'en speed at the apex $z=\L/2$. It is assumed that $Q(R,z)=0$, making Alfv\'en speed continuous at the tube boundary. No restriction is made on $Q$ at the loop ends, $z=0$ and $\L$. Without loss of generality, time may be scaled by setting $a_1=1$ throughout, so the external atmosphere is characterized by one parameter only, $\beta$. 

Although more complex internal densities may be accommodated by the expansion method, for simplicity it will be assumed here that $Q$ takes the form (with $a_1=\L=1$)
\begin{equation}\label{Qbetagamma}
\begin{split}
Q(r,z) &= \frac{1}{a(r,z)^2}-\frac{1}{a_0(z)^2}\\[4pt]
&=\left(1-(r/R)^{1/\alpha}\right)
\left[
(Q_0+1) \rme^{\gamma(1-\sin\pi z)}-\rme^{\beta(1-\sin\pi z)}
\right] = f(r)\,h(z)
,  
\end{split}
\end{equation}
which again represents a continuous Alfv\'en speed at $r=R$. The density scale heights inside and outside the loop match if $\gamma=\beta$, in which case
\begin{equation}
Q(r,z) =  Q_0\,\left(1-(r/R)^{1/\alpha}\right)
\rme^{\beta(1-\sin\pi z)}
,  \label{Qbeta}
\end{equation}
Equation (\ref{Qbetagamma}) corresponds to a loop centre Alfv\'en speed specified by $a(0,z)^{-2}=(Q_0+1) \rme^{\gamma(1-\sin\pi z)}$, representing a hot loop (relative to the external atmosphere) if $\gamma<\beta$, and a cool loop if $\gamma>\beta$.

With Equation (\ref{a0strat}) in place, the eigenvalue problem (\ref{SL}) with zero endpoint conditions is solved numerically for the $L_n$ and $\omega_n$. {For simplicity, attention will be restricted to even modes in $z$ by selecting only expansion functions $w_n(z)$ that are symmetric about $z=\ell/2$. With a symmetric Alfv\'en speed profile, as chosen here, the odd and even modes are decoupled in any case.}


\begin{figure}[tbhp]
\begin{center}
\includegraphics[width=\textwidth]{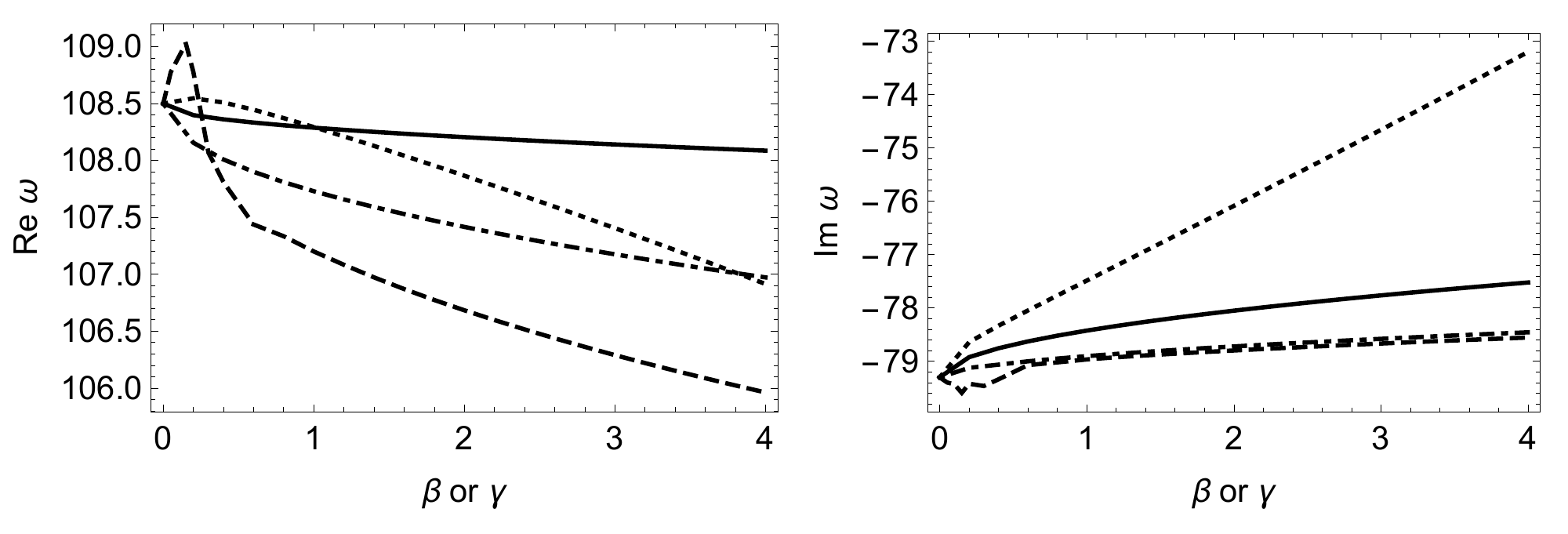}
\caption{ Real (left) and imaginary (right) parts of the fundamental eigenfrequencies for thin loops ($R=0.02$) with $Q_0=0.2$ and $\alpha=0.1$ as functions of $\beta$ or $\gamma$. The full curves correspond to $\gamma=\beta$, i.e., to where the internal and external scale heights are the same. The dashed curves relate to ``hot'' loops where $\gamma=0$ and $\beta$ varies, i.e., the density is uniform in $z$ inside the loop centre, despite the external atmosphere being stratified. The dotted curves correspond to $\beta=0$ with $\gamma$ varying, i.e., an unstratified external atmosphere surrounding a stratified loop. The dot-dashed curves belong to the case $\gamma=\beta/2$, and depict $\omega$ against $\beta$. All eigenfrequencies are calculated with $M=3$, $N=6$, except for the $\gamma=0$ case where $N=12$ for $\beta\le\frac{1}{5}$, and $N=10$ for $\frac{1}{5}<\beta\le1$.}
\label{fig:omBeta}
\end{center}
\end{figure}

\subsection{Stratified Atmosphere Results}
{

For the most part, the fundamental mode is adequately represented with spectral resolution $M=3$, $N=6$ (used throughout unless otherwise noted). That is, there is very little energy in the $m=M$ and $n=N$ last retained expansion functions, as judged by the squared magnitudes of the expansion coefficients $|Y_{mn}|^2$. Indeed, by Parseval's theorem, $ \sum_{m=1}^M \sum_{n=1}^N |Y_{mn}|^2=1$, so the $|Y_{mn}|^2$ are just the fractional energies in each expansion mode. The $Y_{mn}$ are calculated by solving $KAK\,Y=0$ once the frequency $\omega$ has been iteratively adjusted to make $A$ singular (see Eqns.~(\ref{Yeqn}) and (\ref{singular})). The longitudinal expansion modes $w_n$ (solutions of Eqn.~(\ref{SL})) are ordered by decreasing $|L_n^2-\omega^2|$, which assures that the longitudinal external eigenfunctions $w_n$ become increasingly oscillatory as $n$ increases, so that $n=1$ is identifiably the fundamental.

\begin{figure}[tbhp]
\begin{center}
\includegraphics[width=\textwidth]{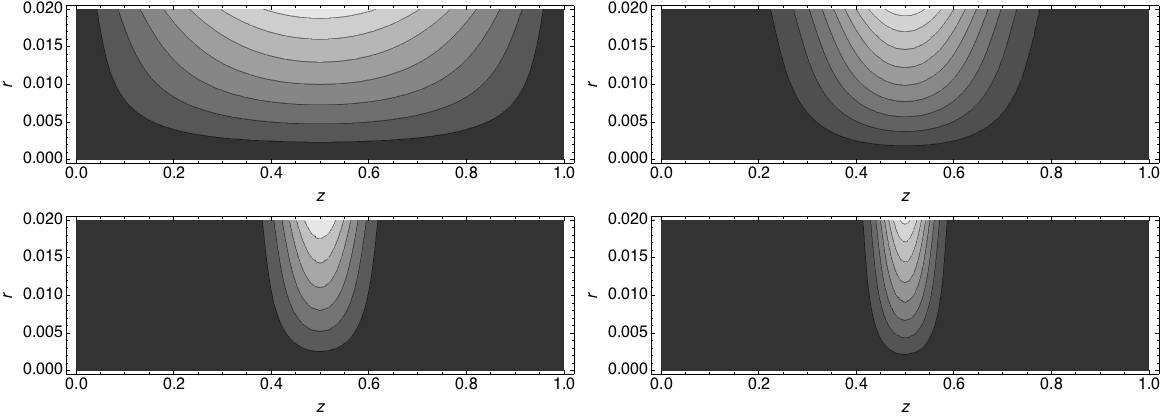}
\caption{The absolute values $|\xi(r,z)|$ of the fundamental displacement eigenfunctions for $R=0.02$, $Q_0=0.2$, and $\alpha=0.1$, at $\beta=\gamma=0$ (top left), 0.05 (top right), 1 (bottom left) and 4 (bottom right). }
\label{fig:ef}
\end{center}
\end{figure}

\begin{figure}[tbhp]
\begin{center}
\includegraphics[width=\textwidth]{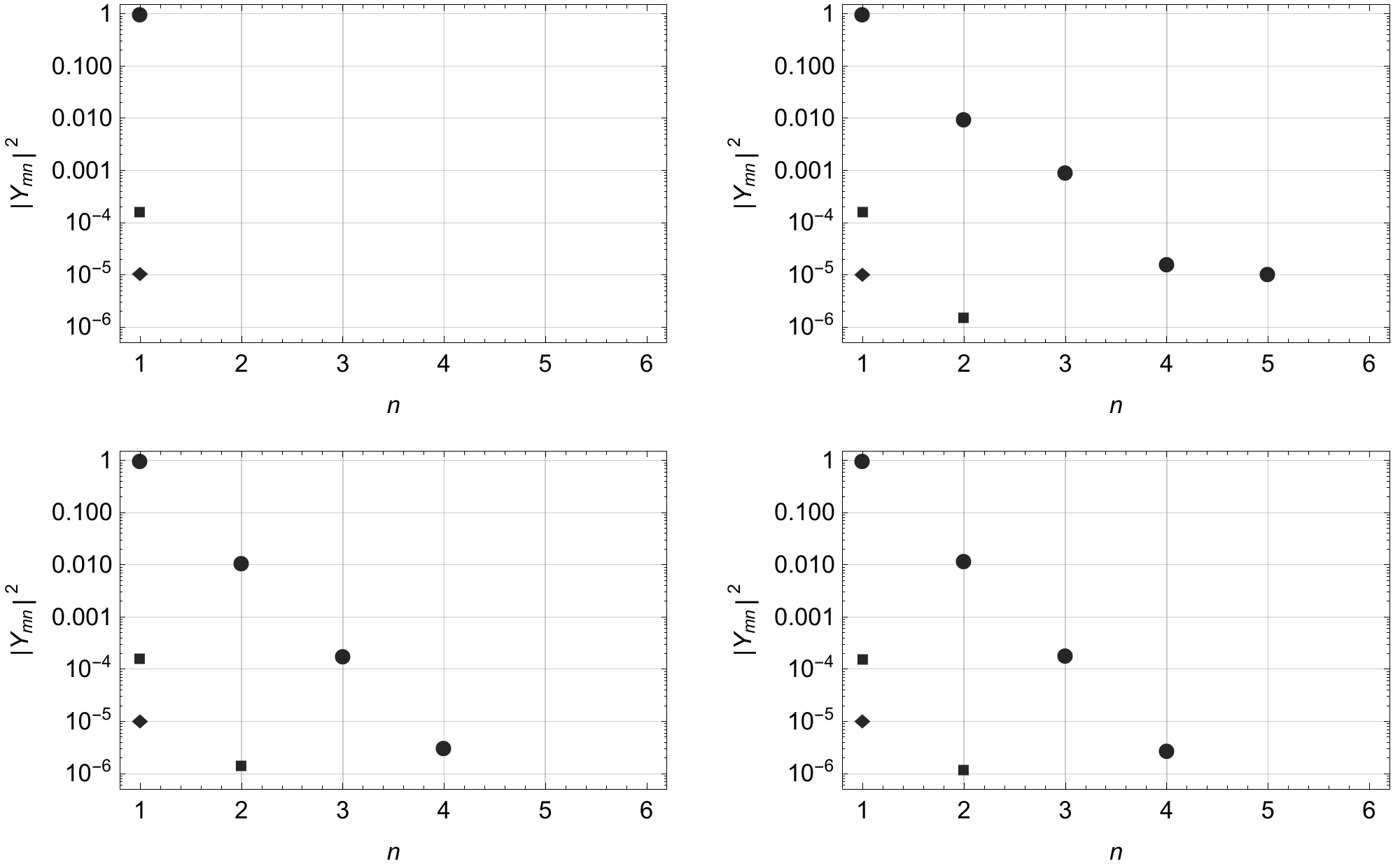}
\caption{The energies $|Y_{mn}|^2$ of the fundamental displacement eigenfunctions of Fig.~\ref{fig:ef} for $m=1$ (circles), $m=2$ (squares), and $m=3$ (diamonds), i.e., $R=0.02$, $Q_0=0.2$, and $\alpha=0.1$, at $\beta=\gamma=0$ (top left), 0.05 (top right), 1 (bottom left) and 4 (bottom right). }
\label{fig:en}
\end{center}
\end{figure}

Since stratification is relevant chiefly to long high loops, attention will be restricted to the ``thin loop'' case, $R=0.02$ (radius divided by length). Figure \ref{fig:omBeta} displays the eigenfrequencies for the case $Q_0=0.2$, $\alpha=0.1$, and varying stratification parameters $\beta$ and $\gamma$. Clearly, these \emph{dimensionless} frequencies dependend only weakly on stratification. 

However, \emph{dimensional} frequency scales as $a_1/\ell$. So, considering a semicircular loop of length $\ell$, fixed base Alfv\'en speed $a_0(0)$, and Alfv\'en (i.e., density) dimensional scale height $H$, the dimensional frequency $\omega_\text{dim}$ scales with the dimensionless frequency $\omega$ according to $\omega_\text{dim}=\omega\,a_0(0)\exp[\ell/(\pi\,H)]/\ell$. Hence, for a highly stratified loop, $\ell\gg H$, the dimensional frequency (and decay rate) increase exponentially with loop length.

Although the dimensionless eigenfrequencies are almost insensitive to stratification, the displacement eigenfunctions $\xi(r,z) = \sum_{m=1}^M \sum_{n=1}^N Y_{mn}\, X_{mn}(r,z)$ evolve considerably as $\beta$ increases (Fig.~\ref{fig:ef}), becoming progressively more compactly situated at the loop apex $z=\ell/2$. This corresponds to the fundamental sausage mode being trapped at the apex by stratification rather than the boundary conditions at $z=0$ and 1. The energies in the constituent expansion functions for each of these eigenfunctions are displayed in Fig.~\ref{fig:en}, showing that $m=n=1$ dominates throughout, as one might expect from Fig.~\ref{fig:ef}.

On the other hand, when the loop itself is unstratified ($\gamma=0$) but the external region is stratified (quantified by $\beta$), some higher overtones become prominent (Fig.~\ref{fig:efg}), ultimately shifting the peak displacement away from the apex, though again it is concentrated by increasing $\beta$. Figure \ref{fig:eng} confirms that the immediate $m=1$ longitudinal overtones are indeed quite prominent. Note that in this case, it was necessary to use enhanced resolution $N=10$ or 12 for small $\beta$.

Figures \ref{fig:en} and \ref{fig:eng} also demonstrate that there is very little contribution of $m>1$ to the fundamental. This is hardly surprising given the wide disparity in different-$m$ eigenfrequencies of the uniform tube exhibited in Table \ref{tab:uniform}.

Finally, very similar behaviour in the eigenfunctions is observed when $\beta=0$ is held fixed (the external atmosphere is unstratified) and $\gamma$ increased (the loop is stratified internally; Fig.~\ref{fig:efi}). In this case the expansion functions $w_n$ are just the $\sin (2n-1)\pi z$, so the compacting of the lobe with increasing $\gamma$ is produced by Fourier superposition, and not a change in $w_1$, as is the predominant mechanism in the other cases.

\begin{figure}[tbhp]
\begin{center}
\includegraphics[width=\textwidth]{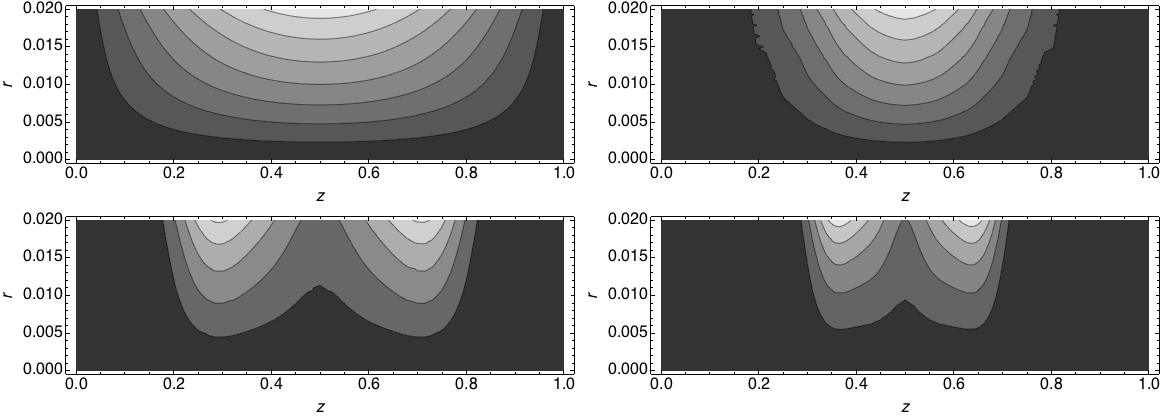}
\caption{The absolute values $|\xi(r,z)|$ of the fundamental displacement eigenfunctions for $R=0.02$, $Q_0=0.2$, and $\alpha=0.1$, with $\gamma=0$ and $\beta=0$ (top left), 0.1 (top right), 1 (bottom left) and 4 (bottom right). Spectral resolution of $M=3$, $N=6$ is used throughout, except for the top right panel where $N=12$.}
\label{fig:efg}
\end{center}
\end{figure}

\begin{figure}[tbhp]
\begin{center}
\includegraphics[width=\textwidth]{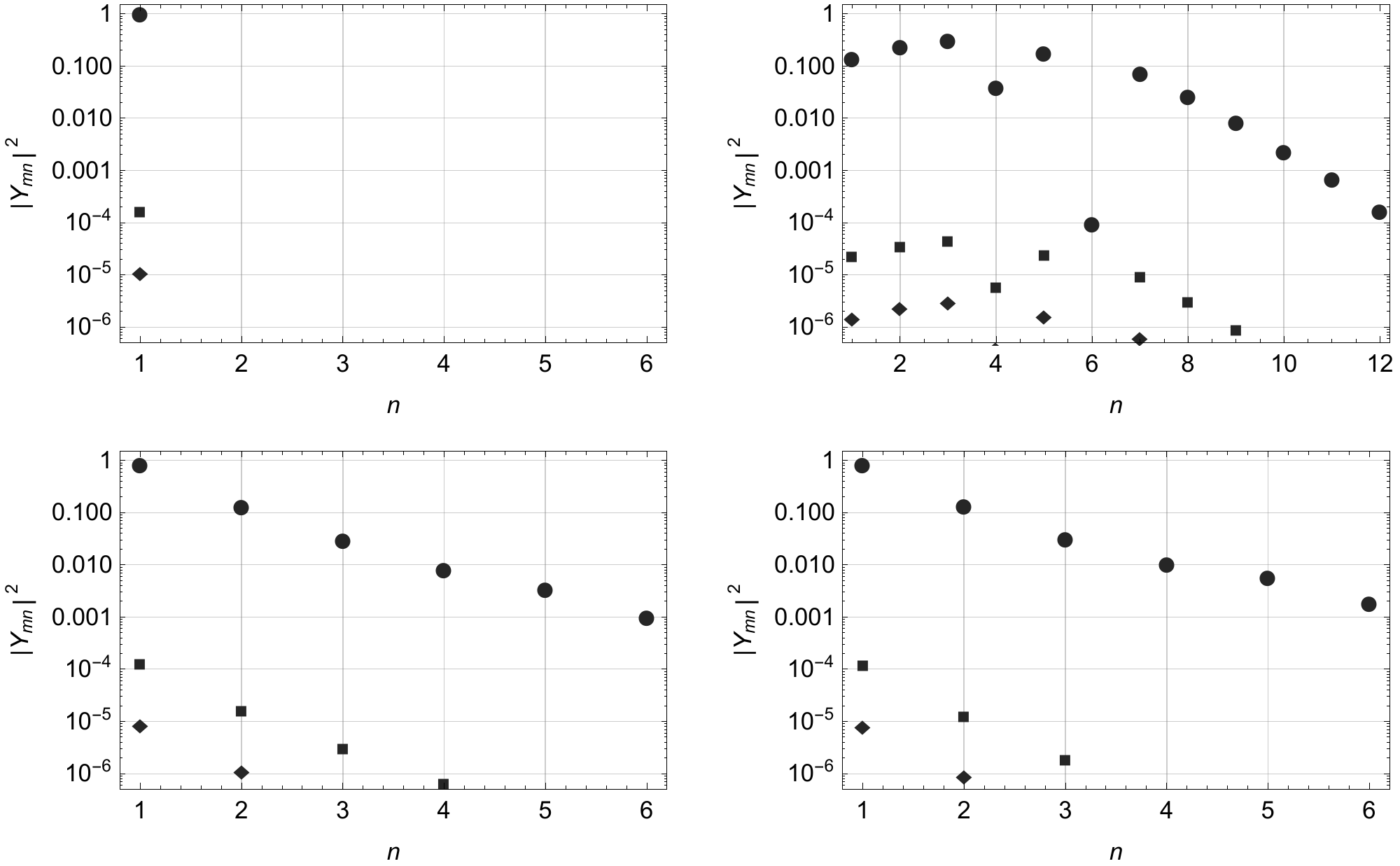}
\caption{The energies $|Y_{mn}|^2$ of the fundamental displacement eigenfunctions of Fig.~\ref{fig:efg} for $m=1$ (circles), $m=2$ (squares), and $m=3$ (diamonds), i.e., $R=0.02$, $Q_0=0.2$, and $\alpha=0.1$, with $\gamma=0$ and $\beta=0$ (top left), 0.1 (top right), 1 (bottom left) and 4 (bottom right). Spectral resolution of $M=3$, $N=6$ is used and adequate throughout, except for the top right panel where $N=12$.}
\label{fig:eng}
\end{center}
\end{figure}

There are two surprising results from this section.
\begin{enumerate}
\item Figure \ref{fig:omBeta} reveals a remarkable insensitivity of the fundamental complex eigenfrequency to stratification, if the Alfv\'en speed $a_1$ at the apex is kept fixed. This suggests that $\omega$ is strongly determined by that local value of $a$.
\item The extent to which the fundamental eigenfunctions are modified by even weak stratification is remarkable. This is a case where the eigenfunction is much more sensitive than the eigenfrequency. This may be understood qualitatively by recourse to eigenvector sensitivity theory in linear algebra, where it is known that eigenvectors depend more sensitively on parameters when the eigenvalues are closely spaced \cite{Golvan96aa}. Reference to Table \ref{tab:uniform} confirms that indeed the eigenvalues of the uniform tube are indeed very closely packed in $n$. This makes nearby harmonics more accessible.
\end{enumerate}

\begin{figure}[tbhp]
\begin{center}
\includegraphics[width=\textwidth]{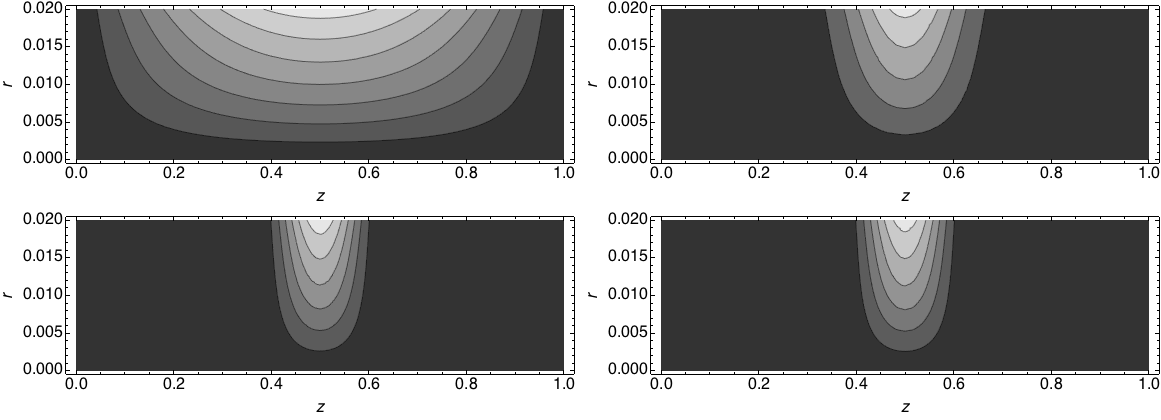}
\caption{The absolute values $|\xi(r,z)|$ of the fundamental displacement eigenfunctions for $R=0.02$, $Q_0=0.2$, and $\alpha=0.1$, with $\beta=0$ and $\gamma=0$ (top left), 0.05 (top right), 1 (bottom left) and 4 (bottom right). }
\label{fig:efi}
\end{center}
\end{figure}

}

\section{Conclusions}

One conceptual and practical advantage of the spectral approach {to the unstratified loop} presented here is that it places the dominant contribution to the eigenvalues on a matrix diagonal, making them almost explicit, with the off-diagonal corrections being moments of the density \emph{inhomogeneity}, vanishing if the tube is uniform. This makes the spectral formulation the most natural for exploring sensitivity of sausage mode eigenfrequencies to density inhomogeneities.

A particular benefit is that the method allows the identification of the linear density kernel $\mathcal{K}_{mn}$, that directly shows the sensitivity of the various modes to density variations in different parts of the loop. Figure \ref{fig:kernels} shows clearly that the $m=n=1$ fundamental is strongly dependent upon the extreme loop edge, and that $m=2$ would be needed to more accurately probe the interior. Observing this mode presents a challenge, but would be worthwhile.

Related to this, it is important to understand that the results presented in Section \ref{resultsIndep} in terms of $\Delta_0$ and $\alpha$ do not actually probe the loop centre, despite $1+\Delta_0$ being ostensibly the density there. That is an artefact of the model (\ref{Delta 1D}). In fact, a density anomaly near $r=0$ is all but invisible to the fundamental mode, so $\Delta_0$ and $\alpha$ should be more properly thought of as parameterizing the boundary layer when used with $m=1$. Higher radial order modes can ameliorate this deficiency somewhat, but the kink mode is a more natural discriminator of loop centre density. 

{Finally, in Section \ref{Sec:strat} the general Sturm-Liouville expansion formalism is applied to allow stratification along the loop, both internally and externally. Numerical results carried out for several cases indicate that stratification leaves dimensionless eigenfrequencies almost unchanged, as measured in a dimensionless system based on the apex external Alfv\'en speed. Of course, stratification will alter this apex Alfv\'en speed and hence the actual dimensional frequency and decay rate. On the other hand, stratification has significant consequences for the fundamental eigenfunction, by either making it more compact around the loop apex as stratification increases, or possibly adding significant contributions of higher longitudinal order ($n$). The main lobe(s) of the fundamental may even be displaced from the apex.}


\vspace{20pt}

\bibliographystyle{iopart-num.bst}
 \input{sausageRev.bbl}



\end{document}

%% file: sausageRev.bbl
\providecommand{\newblock}{}